\documentclass{JHEP3}
\usepackage{epsfig}
\newcommand{\be}{\begin{equation}}
\newcommand{\ee}{\end{equation}}
\newcommand{\bea}{\begin{eqnarray}}
\newcommand{\eea}{\end{eqnarray}}
\newcommand{\IR}{\mathbb{R}} 

\def\IZ{\relax\ifmmode\hbox{Z\kern-.4em Z}\else{Z\kern-.4em Z}\fi}
\newcommand{\IS}{{\bf S}}

\newcommand{\non}{\nonumber \\}

\def\half{{1 \over 2}} 

\def\del{{\partial}}
\def\room{~\rule[-2mm]{0mm}{8mm}}

 \def\hC{\hat{C}} 
\def\ha{{\hat a}} \def\hb{{\hat b}}
\def\hc{{\hat c}} \def\hg{{\hat g}}
\def\hphi{{\hat \phi}}

 \def\co{{\cal O}}
  
\def\cL{{\cal L}} \def\cd{{\cal D}}

\def\talz{\widetilde{\alpha_z}} \def\tb{\widetilde{b}}
\def\ta{\widetilde{a}} \def\tc{\widetilde{c}}
 
\def\al{\alpha} \def\bt{\beta}

\def\lam{\lambda} \def\hxi{{\hat \xi}}

\def\room{~\rule[-2mm]{0mm}{8mm}}
\def\presub{\vspace{.5cm} \noindent}

\def\bi{\begin{itemize}} \def\ei{\end{itemize}}

\def\Schw{Schwarzschild }
\def\({\left(} \def\){\right)}
\def\[{\left[} \def\]{\right]}

\title{ \center{The Power of Action:\\
``The'' Derivation of  the Black Hole Negative Mode}}

\author{
Barak Kol \\
 Racah Institute of Physics\\
 Hebrew University \\
 Jerusalem 91904,
 Israel\\
{\tt barak\_kol@phys.huji.ac.il}}

\abstract{The negative mode of the Schwarzschild black hole is
central to Euclidean quantum gravity around hot flat space and for
the Gregory-Laflamme black string instability. Numerous gauges
were employed in the past to analyze it. Here {\it the} analytic
derivation is found, based on postponing the gauge fixing, on the
power of the action and on decoupling of non-dynamic fields. A
broad-range generalization to perturbations around arbitrary
co-homogeneity 1 geometries is discussed.}

\begin{document}









\section{Introduction}

The freedom to choose coordinates is a central principle of General
Relativity. For a given physical configuration (metric) the theory
produces an unprecedented amount of possibilities for its
description, namely a large gauge symmetry. This gauge symmetry is a
blessing, allowing the use of several coordinate systems each one
adapted to highlight a different feature of the metric. At the same
time it complicates the theory, making it notoriously difficult to
find ``the optimal gauge'' for a given problem. In this paper we
shall solve one outstanding case of this general problem.

The \Schw black hole, a static, spherically symmetric and hence
the simplest of all black holes, is known to possess a single
negative mode discovered by Gross, Perry and Yaffe \cite{GPY}. By
continuity such a mode will appear also for rotating and charged
holes, at least for small angular momentum and charge. While this
mode does not represent a physical instability in the
time-evolution of the black hole (as seen from its precise
definition to be reviewed in the next section) it is responsible
for two important features of black hole physics. First, it plays
a central role in Euclidean Quantum Gravity around hot flat space
\cite{GPY}. The presence of non-zero temperature is known to be
described by a flat space with a periodic Euclidean time
coordinate, whose period is related to the inverse temperature.
The Euclidean black hole has the same asymptotics, and as such
should be included in the path integral as another contributing
saddle point. In particular this additional saddle point
contributes to corrections of the energy density. As explained in
\cite{GPY} the Euclidean black hole solution is related to a
non-perturbative decay of hot flat space (by nucleating a black
hole) and hence the correction to the energy density must have an
imaginary part. On the other hand that correction is proportional
to the square-root of the determinant of fluctuation eigenvalues,
and it is the presence of a single negative mode that guarantees
it is indeed imaginary. The second physical property is the
Gregory-Laflamme (GL) black string instability \cite{GL1} -- a
black-string in 5d was found to be perturbatively unstable to a
non-homogeneous fluctuation if it is ``thin'' enough, namely, as
long as its radius divided by its length (the size of the extra
dimension) is smaller than some critical value. Essentially, the
4d negative mode is the 5d physical instability mode and its
negative eigenvalue is the square of the GL critical value.
Accordingly the negative mode plays a central role in the analysis
of the black-hole black-string phase transition physics associated
with the GL instability \cite{rev}.

Even though the negative mode and its associated eigenvalue is
known already for some 24 years new derivations keep on appearing
\cite{GPY,GL1,Gubser,LargeD,Prestidge,KudohMiyamoto,LG-GL} (see
appendix \ref{previous-gauges}). The new derivations are motivated
by trying to simplify the equations: to decouple the equations for
the various components of the metric perturbation tensor, reducing
them to a single equation (for a ``master field'') and eliminating
unphysical singularities.

One wonders why there are so many different derivations for this
result, and whether we have seen the end of improvements or are
there more surprises waiting for us in this case.

More generally, while a given result may be proven in many ways,
some scientists believe that ``somewhere'' there exists a ``book''
with ``{\it the} proof''. The author believes that the proof below
is ``{\it the} proof'' for the case at hand.

\section{Prudence}
\label{prudence-section}

\noindent {\bf Policy}. In order to optimize our choice of
gauge-fixing our policy in this paper is to postpone any
gauge-fixing as much as possible while collecting all relevant
information for this decision. Therefore in this section we shall be
prudent to define the perturbation eigenvalue problem and to define
the fields, all in a ``pre-gauge-fixing'' form.

\presub {\bf Defining the perturbation eigenvalue problem}. We
would like to consider perturbations around a static spherical
black hole in the background of a flat $d$-dimensional space-time
$\IR^{d-1,1}$. In standard Schwarzschild coordinates
\cite{Schw}\footnote{The generalization to $d \neq 4$ was given in
\cite{Tangherlini}.} it is given by \bea
 ds^2 &=& -f\, dt^2 + f^{-1}\, dr^2 + r^2\, d\Omega^2_{d-2} \non
  f(r) &=& 1- \( \frac{r_0}{r} \)^{d-3} ~, \label{schw-coord} \eea
 where $r_0$, the \Schw radius, is the location of the horizon and
$d\Omega^2_{d-2}$ denotes the metric of the round $d-2$ sphere
$\IS^{d-2}$. For new action-derived coordinates see \cite{nSchw}.

In keeping with our ``pre-gauge-fixing'' policy, we do not employ
the standard definition of the perturbation eigenvalue problem in
terms the Lichnerowicz operator (reviewed below), but rather we
choose a ``pre-gauge-fixing'' formulation as follows. For a general
background $X$ we consider $X'=X \times \IR_z$. In our case $X'$ is
the black string. The space of zero modes of $X'$ is defined as
solutions of the linearized Einstein equations. Due to translation
invariance in the $z$ direction this space of zero modes can be
diagonalized with respect to the $z$-translation operator. We denote
these eigenvalues by $i\, k$ ($k$ is not necessarily real), and we
also use $\lam := -k^2$. The set of values of $\lam$ defines the
\emph{perturbation spectrum of $X$}.

For comparison, an equivalent and more standard definition is in
terms of eigenvalues of the Lichnerowicz operator acting on
perturbations $\delta g_{\mu\nu}$ of the space $X$ \be
 - \cd_\rho \cd^\rho\, \delta g_{\mu \nu} - 2 {R_\mu}^\rho {_\nu}^\sigma\,
\delta g_{\rho \sigma} = \lam \delta g_{\mu \nu} \ee
 (in the notations of \cite{GPY}). Where the perturbations are in
the transverse traceless gauge \bea
 \cd^\mu\, \delta g_{\mu \nu} &=& 0 \non
 \delta g_\mu^\mu &=& 0 ~.\eea
This definition has the advantage that it does not require the
auxiliary $z$ dimension, nor in particular the auxiliary fields
$\delta g_{zz},\, \delta g_{z\mu}$. However, as we shall see, its
gauge choice is restrictive and non-optimal.

\presub {\bf Action approach and fields}. Very generally in
physics it is known that an action, when available, is the most
concise packaging of the equations of motion, and moreover it
enables the widest class of field transformations. Still in
General Relativity (GR) traditionally one writes down the
equations of motion, even though an action principle is
known\footnote{Namely, the Einstein-Hilbert action supplemented in
the case of a fixed boundary by the York-Gibbons-Hawking boundary
term \cite{York,GibbonsHawking}.}. The reason for that is the
large gauge symmetry: in order for the action to encode all the
equations, it must be written as a function of as many fields as
there are equations, while in GR it is common practice to start by
fixing the gauge (an ansatz for the metric) in order to minimize
the number of fields involved and to simplify the equations.

The power of the action formalism leads us to advocate it for GR in
general and specifically for the problem at hand. This strategy
requires working with a maximal set of ``pre-gauge-fixing'' fields
which is also consistent with our above-mentioned policy.

Which fields must be kept in an action approach? For a generic
metric we need to keep all the metric components $g_{\mu\nu}$ in
order to encode all the equations, namely all the components of the
Einstein tensor $G_{\mu\nu}$. However, in the common case of \emph{a
metric with isometries} a reduction in the number of fields is
possible. Given a metric which preserves the isometries, its
Einstein tensor is invariant as well, and in particular some of its
components necessarily vanish. Correspondingly fewer fields are
required.

For the case at hand we seek perturbations of the string which
preserve the $SO(d-1)_\Omega \times U(1)_t \times \IZ_{2,t}$
isometries, namely spherical, stationary and time reflection,
respectively. This \emph{``maximally general ansatz''} is given by
\footnote{Due to the continuous isometries the fields cannot depend
on $t$ nor on the angular coordinates, $\Omega$. The $g_{\Omega i}$
components, where $i \neq \Omega$, must vanish since there is no
spherically symmetric vector field on the sphere, while the $g_{t
j}$ must vanish for all $j \neq t$ due to time reflection combined
with $t$-independence.}  \be
 ds^2 = e^{2\, A}\, dt^2 + e^{2\, B}\, dr^2+ e^{2\, \beta}\, (dz -\al\, dr)^2 + e^{2\, C}\,
 d\Omega_{d-2}^2  \label{ansatz} \ee
which serves to define our notation for the fields. Altogether there
are 5 fields: $A,B,C,\al,\bt$ which are all functions of the $(r,z)$
plane.

We note that our ``maximally general ansatz'' is not the same as
the commonly used term ``the most general ansatz''. The latter
usually means that any metric can be put in that form, while our
ansatz is more general than that: it is constraint-free, namely
all of the Einstein equations can be obtained by varying the
gravitational action with respect to its fields. Actually, if we
fix the $(r,z)$ reparameterization gauge in our ansatz and reduce
the number of fields to 3, it will still be ``most general''.

\presub {\bf Computing the action}. Now we shall derive the
quadratic gravitational action around our black-string background
without gauge fixing either the perturbations, nor the background.
The result is given below in (\ref{quad-action}).

Before computing it, let us discuss a general property which it has.
Generally, when metrics are constrained by isometries, we may
\emph{dimensionally reduce the action} by integrating it over the
``isometry coordinates'' and get a lower dimensional gravitational
action with additional matter content. If moreover we are
considering the action for perturbations around a background with
isometries then we may further reduce along isometries which are
broken by perturbations and get a gauged gravity action where
isometry breaking perturbations are represented by charged fields.
In our case the black string isometries are $U(1)_z \times
SO(d-1)_\Omega \times U(1)_t \times \IZ_{2,t}$ where $U(1)_z$ stands
for $z$-translations and is the only isometry broken by the
considered perturbations. Therefore we expect the action to be
reduced over $t,\Omega,z$ resulting in a $U(1)$ gauged (from $z$),
1d gravity (in $r$) with extra matter.

We write our ansatz (\ref{ansatz}) as \be
 ds^2 = e^{2\, A}\, dt^2 + ds^2_{(r,z)} + e^{2\, C}\,
 d\Omega_{d-2}^2 \non ~. \non \ee
By using the formulae for the Ricci scalar of a fibration (see for
example appendix A in \cite{rev}) we obtain the gravitational action
as \be
 S = \int e^{\Psi}\, dV \( -R_{(r,z)} + K_{ij}\, \del_\mu
 \Phi^i\, \del^\mu \Phi^j - e^{-2\, \hC} \) \ee
where according to our sign conventions $S=-\int R$  (up to
boundary terms), $dV$ is the 2d volume element, $R_{(r,z)}$ is the
2d Ricci scalar, $\mu=r,z$ and we define \bea
 e^{-2\hC} &:=& (d-2)(d-3)\, e^{-2 C} \non
 \Psi &:=& A + (d-2)\, C \non
 \Phi^i &:=& [A,C] \non
 K &:=& -(d-2) \[ \begin{array}{cc}
  0 & 1 \\
  1 & ~~~d-3 \end{array} \] ~.\eea
$\hC$ is a shift of $C$ by a constant to shorten the expressions;
$\Psi$ is a useful expression which appears in $\det(g)$; $A,C$ are
scalars from the 2d point of view and $K$ is their constant kinetic
matrix which can be written also as \be
 K_{ij}\, \Phi^i\, \Phi^j = -(d-2)\, C ( \Psi + A - C) ~.
 \ee

Specifying \be
 ds^2_{(r,z)} = e^{2\, B}\, dr^2+ e^{2\, \beta}\, (dz -\al\, dr)^2
 \non \ee
we obtain \bea
 S &=& \int dr\, dz\, e^{\Psi+B+\bt} \cdot \label{full-action} \\
   &\bigg\{& K_{ij}\, \del_\mu \Phi^i\, \del^\mu \Phi^j - e^{-2\, \hC}
  -2\, e^{-2\, \bt}\, B_z\, \Psi_z - 2\, e^{-2\,
  B}(\bt'+\al_z+\al\, \bt_z)(\Psi'+\al\, \Psi_z) \bigg\} \nonumber  \eea
where a prime $\equiv \del_r$ -- denotes a derivative with respect
to $r$, and a $z$ subscript $\equiv \del_z$.

In our black-string background $\al=\bt=0$ while $A=A_0,B=B_0,C=C_0$
are functions of $r$ only, which depend on the choice of gauge for
the background, and for instance in \Schw coordinates
(\ref{schw-coord}) they are given by \bea
 A_0 &=& \half \log(f) \non
 B_0 &=& -\half\, \log(f) \non
  C_0 &=& \log(r) ~.\eea
Expanding the action (\ref{full-action}) to quadratic order in the
perturbations we obtain \be \fbox{$~~\begin{array}{rcl}
 S_2 &=& \room \int dr\, dz\, e^{\Psi_0+B_0}\, \cL_2  \non
 \cL_2 &=& \room  e^{-2\, B_0} \Big \{  -V_0 \[(\psi+\bt-b)^2+2(\psi+\bt-b)(b-c)+2(b-c)^2\] \non
 &+& \room 2 K_{ij}\, \Phi^i_0\,'\, \( (\psi+\bt-b)\, \phi^j\,' -\al_z\, \phi^j \)
+ K_{ij}\, \phi^i\,'\, \phi^j\,'  \non
  &-& \room  2\,  (\bt'+\al_z) \( \Psi'_0 (\psi+\bt-b) + \psi'\)+2 \ \Psi'_0\, \al_z\, \bt  \Big\} \non
 &+& \room K_{ij}\, \phi_z^i\, \phi_z^j -2\,  \bt_z\, \psi_z \non
 V_0 &:=& \room e^{2\,(B_0-\hC_0)} \label{quad-action}
\end{array}~~~$} ~.\ee
 We have used the background equation of motion $K_{ij}\, \Phi^i_0\,'
\, \Phi^j_0\,'=-V_0$. The $z$ integration above can be carried out
after Fourier expanding all fields, leaving us with a 1d action as
anticipated. Yet we find the current form to be more convenient.

\presub {\bf Gauge invariance}. The quadratic action
(\ref{quad-action}) is invariant under the following gauge
transformations \bea
 \delta a &=& e^{-2 B_0}\, A'_0\, \xi \non
 \delta c &=& e^{-2 B_0}\, C'_0\, \xi \non
 \non
 \delta b &=& e^{-2 B_0}\, \( - B'_0\, \xi + \xi' \) \non
 \delta \al &=& - \hxi'- \xi_z \non
 \delta \bt &=& \hxi_z ~.\label{gauge-trans} \eea
These gauge transformations can be obtained either from the general
form of an infinitesimal diffeomorphism of the background $\delta
g_{\mu\nu} = D_\mu \xi_\nu + D_\nu \xi_\mu$  where $\xi_\mu\,
dx^\mu:=\xi\, dr + \hxi\, dz$ or directly by studying the symmetries
of the action. \footnote{Note that here (and everywhere else) a $z$
subscript $\equiv \del_z$.}

\presub {\bf How to fix the gauge}? Now that the quadratic action
(\ref{quad-action}) is available it may seem that we must fix the
gauge (\ref{gauge-trans}) in order to proceed, namely to supplement
the 5 (linear) equations of motion by two (linear) gauge conditions.
At this point it is not clear how to do that optimally. On the other
hand, the form of the action is not unique and it could still be
transformed. So it makes sense to first transform, simplify and
expose the action's invariant features as much as possible, namely
to find a \emph{canonical form} for it, and only later return to the
issue of gauge fixing.

The action is a quadratic form whose entries are differential
operators.\footnote{More precisely, these are differential operators
in one variable. Actually there are two variables corresponding to
$\del_r$ and $\del_z$ but Fourier decomposition replaces $\del_z \to
i\, k$ and makes it algebraic. This problem is analogous to finding
a canonical form of a quadratic form whose entries are polynomials
(in a single variable).}
 \emph{The set of allowed field transformations}, just like in the Gauss process
for matrices, is the set of invertible transformations, generated by
\be
 \phi^i  \to \phi^i + L \phi^j ~, \ee
where $\phi^i,\, \phi^j,\, j \neq i$ are any two fields and L is any
linear operator, possibly including derivatives. These
transformations are clearly invertible and are also known as
\emph{transformations which keep the measure
invariant}.\footnote{Namely, the Jacobian for the transformation of
the fields is 1.} In the next section we proceed to simplify the
action through the use of these transformations.

\section{The Power of Action}
\label{power-section}

In this section we shall reap the fruit of our prudence to adopt the
pre-gauge-fixing policy and the power of the action will be
demonstrated: by using the invertible field transformations we will
succeed to decouple parts of the action.

In the presence of a gauge symmetry the kinetic part of the action
is necessarily degenerate, since due to the gauge freedom not all of
the fields are determined by initial conditions. The fields which do
not appear in the kinetic term are called \emph{``non-dynamic''
fields}, and they are a good place to start looking.

\presub {\bf Decoupling the non-dynamic fields}. Inspecting the
quadratic action (\ref{quad-action}) one notices that the
non-dynamic fields (with respect to $r$) are $\al,\, b$ (a vector
and tensor from the $r$ perspective), that is, they do not appear
in $\co\(\del_r^2\)$ terms. The part of the Lagrangian density
which involves the non-dynamic fields is \be
 \cL_{2} \supset  e^{-2\, B_0}
 \left\{ \begin{array}{c} \[ \begin{array}{cc} \al_z & b \end{array} \] \\ \hspace{.1cm} \end{array}
  L_{ND} \[\begin{array}{c} \al_z \\ b
 \end{array}\]
 - 2\, \begin{array}{c} \[ \begin{array}{cc} L_\al & L_b \end{array} \] \\ \hspace{.1cm} \end{array}
 \[\begin{array}{c} \al_z \\ b \end{array} \] \right\}
 \label{nd-quad-action} \ee where \bea
 L_{ND} &:=& \[ \begin{array}{cc}
  0                     &  \Psi'_0 \\
   \Psi'_0  & -V_0  \end{array} \] \\ \label{LND} \non
  L_\al &:=&  K_{ij} \Phi^i_0\,'\, \phi^j + \Psi'_0\, \psi +
  \psi'  \non
  L_b &:=& -V_0\, c +
   K_{ij} \Phi^i_0\,'\, \phi^j\,' - \Psi'_0\, \bt'  - e^{+2 B_0}\,
   \psi_{zz} \label{Lalb}
  \eea
By shifting the $\al_z,b$ according to \be
 \[ \begin{array}{c} \talz \\ \tb \end{array} \] :=
 \[ \begin{array}{c} \al_z \\ b \end{array} \]
- L_{ND}^{~-1}\, \[ \begin{array}{c} L_\al \\ L_b \end{array} \]
\label{def-tND} \ee
 where \be
 L_{ND}^{~-1} = \frac{1}{{\Psi'_0}^2}\[ \begin{array}{cc}
  V_0                     &  \Psi'_0 \\
   \Psi'_0  & 0  \end{array} \] \label{LND-inverse} \ee
 the action decouples. The non-dynamic part is simply the original
non-dynamic part with $[\al_z,\, b] \to [\talz,\, \tb]$, namely \be
 \fbox{~\rule[-4mm]{0mm}{12mm} $~~
S_{ND} = \int dr\, dz\, e^{\Psi_0-B_0}
 \left\{ \begin{array}{c} \[ \begin{array}{cc} \talz & \tb \end{array} \] \\ \hspace{.1cm} \end{array}
  L_{ND} \[\begin{array}{c} \talz \\ \tb \end{array}\] \right\}
   ~~$} \label{def-SND} \ee
 and the dynamic part is supplemented by
\be \Delta \cL_D :=  - e^{-2\, B_0} \cdot
  \begin{array}{c} \[ \begin{array}{cc} L_\al & L_b \end{array} \] \\ \hspace{.1cm} \end{array}
  L_{ND}^{~-1} \[\begin{array}{c} L_\al \\ L_b \end{array}\]
   \label{DeltaSD} \ee

The equations of motion for $(\talz,\, \tb)$ derived from $S_{ND}$
(\ref{def-SND}) are simply \be
 0=\talz =\tb \label{eom-talz-tb} ~.\ee

Notice that the existence of an inverse for $L_{ND}$
(\ref{LND-inverse}) was essential for this process, and this in turn
depended on the fact that the non-dynamic action $L_{ND}$
(\ref{LND}) was algebraic (while two derivative terms are forbidden
by the definition of non-dynamic fields, one derivative terms are
not). This process is equivalent to \emph{``integrating out'' the
non-dynamic fields}, namely to solve their equations of motion, and
substitute the solutions back into the action. Also, while our shift
(\ref{def-tND}) involves differential operators, the action still
does not contain terms with more than two derivatives due to the
non-dynamic nature of the shifted fields.

Altogether we achieved a decoupling of the action into non-dynamic
fields $\talz,\, \tb$ and dynamic fields $a,c,\bt$. The quadratic
action (\ref{quad-action}) is given now by \bea
 S_2 &=& S_{ND} + \int
dr\, dz\, e^{\Psi_0+B_0}\, \cL_D  \non
 \cL_D &:=& e^{-2\, B_0} \Big\{  -V_0 \[(\psi+\bt)^2-2c\, (\psi+\bt)+2\, c^2\] \non
 &+& 2 \, (\psi+\bt)\, K_{ij}\, \Phi^i_0\,'\,   \phi^j\,'
 +  K_{ij}\, \phi^i\,'\, \phi^j\,' \non
 &-&  2\,  \bt'\, \( \Psi'_0 (\psi+\bt) + \psi'\) \Big\} \non
 &+& K_{ij}\, \phi_z^i\, \phi_z^j + \Delta \cL_D  \label{quad-action-D+ND} \eea
We now proceed to simplify $\cL_D$.

\presub {\bf Gauge invariance}. Performing the gauge variations
(\ref{gauge-trans}) on the tilded variables (\ref{def-tND}) one
finds that \emph{$(\talz,\tb)$ are gauge invariant}. This is not
surprising since once the action decouples we expect the gauge
transformations to affect only one of its parts and anyway an
algebraic action does not offer any symmetry that could be gauged.

The gauge symmetry reduces to \bea
 \delta a &=& e^{-2 B_0}\, A'_0\, \xi \non
 \delta c &=& e^{-2 B_0}\, C'_0\, \xi \non
 \delta \bt &=& i\, k\, \hxi ~.\label{gauge-trans-D} \eea
Since $\delta \bt \propto \hxi$ (for $ 0 \neq i\, k \equiv \del_z$)
the action (\ref{quad-action-D+ND}) must be independent of $\bt$, as
can be confirmed by direct calculation.\footnote{Gauge invariance
implies that $\delta S/\delta \hxi$ is a total derivative but by
(\ref{gauge-trans-D}) $\delta S/\delta \hxi \propto \delta S/\delta
\bt$ and hence the equation of motion for $\bt$ vanishes.}

Similarly the action can depend only on the gauge invariant
combination of $a,c$, namely the Wronskian-like expression $C'_0\, a
- A'_0\, c$. This combination is unique up to multiplication by a
function of $r$, and inspection of the kinetic term suggests to
define the dynamic gauge-invariant field to be \be
 \tc:=\sqrt{(d-2)(d-1)}\, \( c -\frac{C'_0}{\Psi'_0}\, \psi \) \label{def-tc}
\ee

\presub {\bf The dynamic part of the action.} In terms of the gauge
invariant field $\tc$ the dynamic part of the action is given by
 \be \fbox{ $~~\begin{array}{rcl}
  S_D &=& \room \int dr\, dz\, e^{\Psi_0+B_0}
 \left\{ e^{-2 B_0}\, \tc\,'\,^{2} + \tc_z^{~2}+V(r)\, \tc^{~2} \right\} =
 \non
&=& \room \int dr\, dz\, e^{\Psi_0+B_0}
 \left\{ (\del \tc)^2 + V(r)\, \tc^{~2} \right\} \label{def-SD}
 \end{array}~~$} \ee
 where the ``potential'' is
 \be \fbox{\room $~~
 V(r):=-2(d-1)(d-3)\, e^{-2 C_0}\, \frac{A_0^{'2}}{\Psi_0^{'2}} \label{def-V}
 ~~$}~~.
  \ee
The equation of motion for $\tc$ is \be
 -\triangle \tc + V(r)\, \tc = -k^2\, \tc \label{eigenvalue} \ee
 where $\triangle:=e^{-\Psi_0-B_0}\, \del_r  e^{\Psi_0-B_0}
\del_r$ is the Laplacian in the black hole background. The analogy
with a Schr\"{o}dinger eigenvalue problem motivated the notation
$V(r)$.

\begin{figure}[t!] \centering \noindent
\includegraphics[width=7cm]{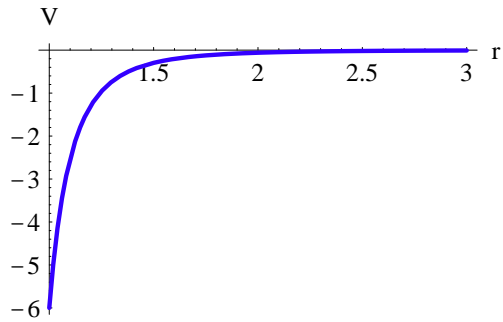}
\caption[]{A graph of the perturbation potential $V$ as a function
of the radial coordinate $r$ (\ref{Vschw}) in $r_0=1$ units. Drawn
at $d=4$ -- in higher dimensions $V$ becomes even more concentrated
around $r/r_0=1$.}\label{V-figure}
\end{figure}

\presub {\bf Properties of $V(r)$.} The potential $V(r)$ is
manifestly independent of the gauge for the background. In \Schw
coordinates (\ref{schw-coord}) it is given by \bea
 V(r) &:=& -2(d-1)(d-3)\, \frac{f'^2}{(2(d-2)\,f+r\, f')^2} =\non
 &=& -\frac{2(d-1)(d-3)^3}{r^2 \Big( 2(d-2) (r/r_0)^{d-3} -(d-1) \Big)^2} \label{Vschw} ~.
 \eea
This is precisely the form given in \cite{LG-GL} eq. (3.6). It can
be seen that $V(r)$ is a negative potential concentrated strongly
near the horizon: it is finite on the horizon and it vanishes
asymptotically. See figure \ref{V-figure} for a graph of $V$.

\subsection{Summary - the negative mode}

While transforming the action we were able to completely eliminate
the gauge, thereby \emph{rendering the gauge-fixing question
obsolete}. The final gauge invariant action is given by \be
 S_2 = S_D[\tc] + S_{ND}[\talz,\tb] \label{canonical-action} \ee
where the dynamic part is given by (\ref{def-SD}) and the
non-dynamic part by (\ref{def-SND}).

The spectrum of perturbations around the black hole background is
given by the eigenvalue problem (\ref{eigenvalue}) with $V(r)$
defined in (\ref{def-V}). That problem has a single negative
eigenvalue which defines the Gross-Perry-Yaffe eigenvalue
$\lam_{GPY} \equiv k_{GL}^2$, where $k_{GL}$ is the Gregory-Laflamme
critical wavenumber.

Our canonical form for the action (\ref{canonical-action}) is so
simple and its derivation so straightforward that it is hard to
imagine a ``simpler'' form. It is certainly as simple as any of the
previous forms (see appendix \ref{previous-gauges}). It contains a
single-field master equation with no singularities between the
horizon and infinity, and the non-dynamic fields are determined
algebraically. It turns out that our eigenvalue equation
(\ref{eigenvalue}) is equivalent to the one already derived in
\cite{LG-GL}. Still the picture now is clearer: we know not to
expect a simpler expression; the non-dynamic fields are algebraic,
contrary to \cite{LG-GL}; the nature of the master field is
clarified to be a gauge invariant combination of the ``scalars''
$a,c$ ; the potential is clarified to originate from both the sphere
potential $\sim e^{-2 C}$ and from the potential $\Delta \cL_D$
(\ref{DeltaSD}) which arose during the decoupling (integrating out)
of the non-dynamic fields. In addition, in the next section we will
present a single field master equation for the $k=0$ sector, which
improves on the 2-field equation of \cite{LG-GL}.

\section{The $k=0$ case}

The $k=0$ zero-mode sector of perturbations around the black string
background provides another example for our techniques.

Considering the quadratic action (\ref{quad-action}) we notice that
$\al$ appears only in the combination $\al_z$ and therefore $\al$
does not participate in the $k=0$ sector. Moreover the gauge
transformation parameter $\hxi$ appears in the transformation
(\ref{gauge-trans}) only as $\hxi_z$ (once one operates with
$\del_z$ on the equation for $\delta \al$) and therefore neither
does $\hxi$ participate in this sector.

Altogether there are 4 fields in this sector: $a,b,c$ and $\bt$ and
one gauge function $\xi$. From our experience with $k \neq 0$ we
expect that the gauge function will eliminate one field, and one
will be non-dynamic leaving us with two dynamic fields. Would it be
possible to decouple them?

\presub {\bf Weyl rescaling}. The answer is actually known. The
$k=0$ sector is the same as a Kaluza-Klein reduction over the $z$
direction. In this case it is known that the $d+1$ dimensional
metric field decouples into a scalar related to $g_{zz}$ and the $d$
dimensional metric. The decoupling is achieved by an appropriate
Weyl-rescaling. In practice we define the Weyl rescaled $d$-metric
$\hat{g}_{\mu\nu}$ by \be
 ds^2 = e^{2 \gamma}\, \widehat{ds}^2 + e^{2 \bt}\, dz^2 \ee
 where the conformal factor is determined by \be
 \gamma =-\frac{1}{d-2}\, \bt ~.\ee
 Altogether the effect on the fields is \be
 \[ \begin{array}{c} a \\ b \\ c \end{array} \] =
 \[ \begin{array}{c} \ha \\ \hb \\ \hc \end{array} \] -
 \frac{\bt}{d-2} \[ \begin{array}{c} 1 \\ 1 \\ 1 \end{array} \] ~. \label{Weyl-rescaled} \ee

Once we substitute the Weyl-rescaled fields (\ref{Weyl-rescaled})
into the $k=0$ sector of the quadratic action (\ref{quad-action})
$\bt$ decouples. Its part in the action is \be
 \fbox{\room $~~
 S_{\bt} = \int dr\, dz\, e^{\Psi_0-B_0}\, \frac{d-1}{d-2}~
 \bt\,'^2
 ~~$} \label{Sbt} \ee
 which represent a minimally coupled scalar, as expected, and
from now on all fields should be understood to have $k=0$ and all
actions are restricted to this sector. Note that $\bt$ is gauge
invariant for $k=0$ (\ref{gauge-trans}).

\presub {\bf Non-dynamic action}. The rest of the action contains
the fields $\ha,\hb,\hc$. Following our successful procedure we
locate the non-dynamic fields and attempt to decouple them. Here
$\hb$ (or equivalently $b$) is the only non-dynamic field and the
part of the action where it appears is \be
 \cL_{2} \supset  e^{-2\, B_0}\,
 \left\{-e^{2 B_0-2 \hC_0}\, \hb^2
 -2\,  L_b\, \hb  \right\}
 \label{nd-quad-action-k0} \ee
where from (\ref{Lalb}) \be
 L_b :=    K_{ij} \Phi^i_0\,'\, \hphi^j\,'  -V_0\, \hc ~.  \ee
 The shifted $b$ is given by \be
 \tb := \hb + \frac{L_b}{V_0} = \hb-\hc +\frac{K_{ij} \Phi^i_0\,'\, \hphi^j\,'}{V_0} \ee
 and is gauge-invariant as before.
The non-dynamic sector of the $k=0$ action is given by \be
\fbox{\room $~~
 S_{ND} = \int dr\, dz\, e^{\Psi_0-B_0}
 \left\{-V_0 ~ \tb^2 \right\}
 ~~$}
 \label{def-SND-k0} ~,\ee
 the corresponding equation of motion is \be
 \tb =0 ~,\ee
and the added term to the dynamic action is \be
 \Delta \cL_D := +e^{-2\, B_0}\, \frac{1}{V_0}~ L_b^{~2}~~.
  \label{DeltaSD-k0} \ee

\presub {\bf Eliminating the gauge and final decoupled form}. We are
left with the dynamic action, which is a function of $\ha,\hc$ and
is invariant under the gauge function $\xi$. As for $k \neq 0$ there
is a single gauge invariant combination up to multiplication by a
function of $r$ and here it is convenient to define \be
 \ta := \sqrt{\frac{d-2}{d-3}}\, \( \ha - \frac{A'_0}{G'_0}\, \hg \)
 \label{def-ta-k0} \ee
 where it is useful to define \be
 G := A + (d-3)\, C \ee
 and accordingly the decomposition into background and fluctuations
is $G=G_0+g$ and Weyl-rescaling gives $g=\hg-\bt$. In terms of $A,G$
the kinetic matrix is $K_{ij}\, \Phi^i\,
\Phi^j=(d-2)/(d-3)\,(A^2-G^2)$. The remaining dynamic action is a
function of $\ta$ alone and it reads \be \fbox{\room $~~
 S_{\ta} = \int dr\, dz\, e^{\Psi_0-B_0}\,
 \frac{G'_0\,^2}{G'_0\,^2-A'_0\,^2}~ \ta\,'^{~2}
 ~~$} \label{def-Sta-k0}\ee
 where the constant prefactor in the definition of $\ta$
(\ref{def-ta-k0}) was chosen as to eliminate a multiplicative
constant from the action. This is an action of a scalar field with a
($r$ dependent) non-standard Kinetic term. Note that
$G'_0\,^2-A'_0\,^2=(d-3)/(d-2)\, V_0$. In \Schw coordinates
(\ref{schw-coord}) this equation reads \be
 S_{\ta} = \int dr\, dz\, r\, \( r/r_0\)^{d-3}\, \(1-\half
 \(\frac{r_0}{r}\)^{d-3} \)^2 ~ \ta\,'^{~2} ~.
\ee This formula was tested to reproduce the $k=0$ zero mode of the
black hole which corresponds to varying $r_0$.

Is summary, the total gauge invariant action in the $k=0$ sector is
given by \be
 S_2 = S[\bt] + S[\ta] + S_{ND}[\tb] \label{canonical-action-k0} \ee
where the three summands are defined by
(\ref{Sbt},\ref{def-Sta-k0},\ref{def-SND-k0}), respectively.
Together with the decoupled form of the $k \neq 0$ sector
(\ref{canonical-action}) we have completely decoupled the quadratic
action in the background of the \Schw black hole.

\subsection{Generalizations}

\noindent{\bf Generalization}. Now that we have seen the workings of
this method in two cases, we may draw some general lessons. We saw
that each gauge function is responsible for a non-dynamic field. The
action in the non-dynamic sector is purely algebraic. The
non-dynamic sector can be decoupled from the dynamic sector after a
redefinition of the non-dynamical fields which renders them
gauge-invariant. The remaining dynamical fields depend algebraically
on the gauge functions and thus each gauge function eliminates a
field, leaving a gauge-invariant dynamical action.

How general is this procedure? The essential property was that the
fields depended essentially on a single variable, which was $r$ in
our case. \emph{I claim that this method should work very generally
in any essentially 1d case, including perturbations of any
co-homogeneity 1 metric (or gauge field)}. In particular, in the
presence of $n_F$ fields and $n_G$ gauge functions there will be a
canonical form for the action which will include an algebraic sector
with $n_G$ gauge invariant fields, and a dynamic sector with
$n_F-2\, n_G$ gauge invariant fields. The precise domain of validity
of this statement is under study.

\presub {\bf Short cut}. Once we know that all dependence on the
gauge-variant fields must vanish, we may leave behind our
``pre-gauge-fixing'' policy and equate those fields to zero from the
outset. For example in the $k \neq 0$ case we may start by writing
the action as a function of 3 fields only: $\al,b,c$ ($c$ can be
replaced by any linear combination of $a,c$). Then one proceeds as
before to decouple $\al,b$ through redefinition. One could say
therefore that the ``optimal gauge'' for this problem is $0=\bt=a$
(in the sense just described -- in particular $0=\bt=c$ is as
``optimal'').

\presub {\bf Spin-offs}. Let us stress some general lessons learned
while solving the problem of perturbations around the \Schw black
hole. We saw that the power of the action formalism does not fail
when entering General Relativity. We saw that for the action to
encode all equations it is important to adopt the ``pre-gauge-fixing
policy'', and use the ``maximally general'' ansatz for a given
isometry group. In particular we stress the alternative formulation
of the Lichnerowicz operator in terms of an extra dimension. We
defined ``allowed field transformation'' within the action formalism
to be invertible transformations, even if differential, and used
them to transform the action into a ``canonical form''.

\presub {\bf Open directions}. Finally I would like to mention some
open directions. In this paper we worked at quadratic order but this
method should certainly work beyond it. A more difficult question is
whether any of these properties continues to hold for systems with
more essential dimensions, such as the essentially 2d ($r,\theta)$
Kerr black hole.

\vspace{0.5cm} \noindent {\bf Acknowledgements}

I would like to thank Evgeny Sorkin, Vadim Asnin and Michael Smolkin
for collaboration on related projects, and Ofer Aharony for a
discussion.

This research is supported in part by The Israel Science Foundation
grant no 607/05 and by the Binational Science Foundation
BSF-2004117.

\appendix
\section{Previous gauges}
\label{previous-gauges}

In this section we list in chronological order some of the gauge
choices which were previously used to find the negative mode.

\presub {\bf Gross-Perry-Yaffe (1982)} \cite{GPY}. In $d=4$, the
transverse ($D^\mu h_{\mu\nu}=0$) traceless\footnote{Although in
principle we are allowed only one gauge condition here ($r$
reparameterization) tracelessness
 comes along with transversality.} ($a+b+2c=0$)
Lichnerowicz ($\al=\bt=0$) gauge was employed and a master equation
 was obtained (their 5.21), which in our notation reads \be
 -f\, b'' - \frac{8\, r^2-22 r_0\, r + 12 r_0^2}{r^2(2r-3r_0)}\, b
  +\frac{8 r_0}{r^2(2r - 3 r_0)}\, b = \lam\, b ~.\ee
Note that this equation has an unphysical singularity at $r=3
r_0/2$, the radius of the closed light-like geodesic.

\presub {\bf Gregory-Laflamme (1993)} \cite{GL1}. Here not only
marginally tachyonic modes were sought but also time-dependent modes
$\propto e^{\Omega t}$. The Lichnerowicz gauge $\al=\bt=0$ was used,
and a master equation was obtained (their eq. 10) for all $d$. The
master field is $h^{tr}$ and since the equation is 4 lines long it
is not reproduced here.

\presub {\bf Gubser (2001)} \cite{Gubser}. Worked in $d=4$ with a
conformal-type gauge (in the $(r,z)$ plane) \be
 ds^2=f\, e^{2 a}\, dt^2 + e^{2 \bar{b}}\, \(f^{-1}\, dr^2 +
 dz^2 \) + r^2 e^{2 c}\, d\Omega_2\;^2 ~.\ee
 An eigenvalue problem for 2 fields was obtained as follows \bea
 -2r(r-1)\, a''+(4r-1)\, a'-2 c' &=&2\, \lam\, r^2\, a \non
 -r(r-1)(4r-3)\, c'' - (8 r^2-16 r +9) c' - 2\, c +&& \non
 +3(r-1)\, a' + 2\, a &=& \lam\, (4r-3)r^2\, c \eea
 in units where $r_0=1$.

\presub {\bf Sorkin-Kol (2004)} \cite{LargeD}. Generalized the
Gross-Perry-Yaffe computation to arbitrary $d$ with essentially the
same gauge. The same computation was carried out in \cite{Prestidge}
only the expressions in the master equation were not simplified
enough. The obtained master equation is \bea
 -f\, b'' + \frac{2r^2(f f'' - f^{'2})-r(d-2)f f' + 2d f^2}{r(r
 f'-2f)}\, b' + && \non
 +\frac{r^2 f' f'' r[2(d-1)f f'' - (d+2) f^{'2}]+4 f f'}{r(r f'-2 f)} \, b &=& \lam\, b
 \eea
 Just like \cite{GPY} it suffers from an unphysical singularity.

\presub {\bf Kudoh-Miyamoto (2005)} \cite{KudohMiyamoto}. With the
help of the Harmark-Obers coordinates \cite{HOcoord} a master
equation was obtained \be
 -f\, \hat{b}'' -\frac{(d-1)f^2-(d-3)(3d-8)f+(d-3)^2}{r \(d-3+(d-1)f\)}\,
 \hat{b}' -\frac{2(d-3)^2(1-f)}{r^2 \(d-3+(d-1)f\)}\, \hat{b}  = \lam\, \hat{b}
  \ee
where $\hat{b}$ is defined within the Harmark-Obers coordinates
and the notation is unrelated to any previous use of $\hat{b}$.
This is a single field master equation without any unphysical
singularities between the horizon and infinity. However, the
expressions are more complicated than our final master equation
(\ref{eigenvalue}).

\presub {\bf Sorkin-Kol (2006)} \cite{LG-GL}. Here the chosen
gauge was $\al=0$ together with a choice of $b$ such that the
$\al$-constraint $\delta S/\delta \al=0$ is proportional to
$\psi'$ (and there is no term without derivatives). The master
field is $a$ and the master equation is the same as the one
derived here (\ref{eigenvalue},\ref{Vschw}) \be
 -\frac{1}{r^{d-2}}\del_r \( f r^{d-2}\, a' \)
 - \frac{2(d-1)(d-3)\, f'^2}{(2(d-2)\,f+r\, f')^2}\, a= \lam\, a
 ~.
\ee

\end{document}